\documentclass{emulateapj}

\slugcomment{To be appeared in ApJL}

\shorttitle{Distributions of Flares with and without CMEs}
\shortauthors{Yashiro et~al.}

\begin{document}

\title{Different Power-law Indices in the Frequency Distributions of
Flares with and without Coronal Mass Ejections}

\author{S. Yashiro,\altaffilmark{1,2} S. Akiyama,\altaffilmark{1,2}
N. Gopalswamy,\altaffilmark{2} and R. A. Howard\altaffilmark{3}}

\altaffiltext{1}{Catholic University of America, Washington, DC 20064}

\altaffiltext{2}{NASA Goddard Space Flight Center, Greenbelt, MD 20771}

\altaffiltext{3}{Naval Research Laboratory, Washington, DC 20375}

\begin{abstract}

We investigated the frequency distributions of flares with and without
coronal mass ejections (CMEs) as a function of flare parameters (peak
flux, fluence, and duration of soft X-ray flares). We used CMEs
observed by the Large Angle and Spectrometric Coronagraph (LASCO) on
board the Solar and Heliospheric Observatory (SOHO) mission and soft
X-ray flares (C3.2 and above) observed by the GOES satellites during
1996 to 2005.  We found that the distributions obey a power-law of the
form: $dN/dX \propto X^{-\alpha}$, where $X$ is a flare parameter and
$dN$ is the number of events recorded within the interval [$X$,
$X+dX$]. For the flares with (without) CMEs, we obtained the power-law
index $\alpha=1.98\pm0.05$ ($\alpha=2.52\pm0.03$) for the peak flux,
$\alpha=1.79\pm0.05$ ($\alpha= 2.47 \pm 0.11$) for the fluence, and
$\alpha=2.49\pm0.11$ ($\alpha=3.22\pm0.15$) for the duration. The
power-law indices for flares without CMEs are steeper than those for
flares with CMEs. The larger power-law index for flares without CMEs
supports the possibility that nanoflares contribute to coronal
heating.

\end{abstract}

\keywords{Sun: flares --- Sun: CMEs --- Sun: corona}

\section{Introduction}

Heating of the solar corona is one of the fundamental problems in
solar physics. Solar flares have been proposed as a heat source, but
the observed flares do not supply enough energy to keep the coronal
temperature at million degrees. However, tiny flares known as
nanoflares, whose intensity is below the observational limits may be
able to heat the corona \citep{parke88}. Since the nanoflares cannot
be detected as discreet events with the current observational
capability, their occurrence frequency distribution is often
extrapolated from the observed flares. The flare frequency
distributions can be represented by a power-law of the form:
$dN/dE~\propto~E^{-\alpha}$, where $E$ is flare energy and $dN$ is the
number of events recorded within the interval [$E$,~$E+dE$]. When
$\alpha < 2$, only larger flares dominantly contribute to coronal
heating \citep{hudso91}, meaning that nanoflares cannot
contribute. Flare peak flux or peak count rate have been used to
obtain the power-law index since it is difficult to measure the total
flare energy. Many authors have examined $\alpha$ for various
parameters of flares and flare-related phenomena, and found to be
smaller than 2 \citep[e.g.,][and references
therein]{crosb93,aschw98}. The only exception was for quiet-region
flares observed in EUV \citep[$\alpha=2.3-2.6$;][]{kruck98}.

After the discovery of CMEs in 1971, the relation between flares and
CMEs have been studied extensively (see \citealt{kahle92} for
review). A close relation is also indicated from the similarity
between the derivative of the X-ray light curve and CME acceleration
profile \citep{zhang01,vrsna04}. However, not all flares are
associated with CMEs. Even X-class flares (about 10\% of them) lack
CME association \citep{yashi05}. Since a large, uniform and extended
data base on CMEs has become available for the first time from SOHO,
we can perform an extensive statistical analysis for a detailed
examination of flares with and without CMEs. In this paper, we show
the frequency distributions for flares with and without CMEs, and
discuss their implications for the problems of coronal heating.

\begin{figure*}
\epsscale{0.9}
\plotone{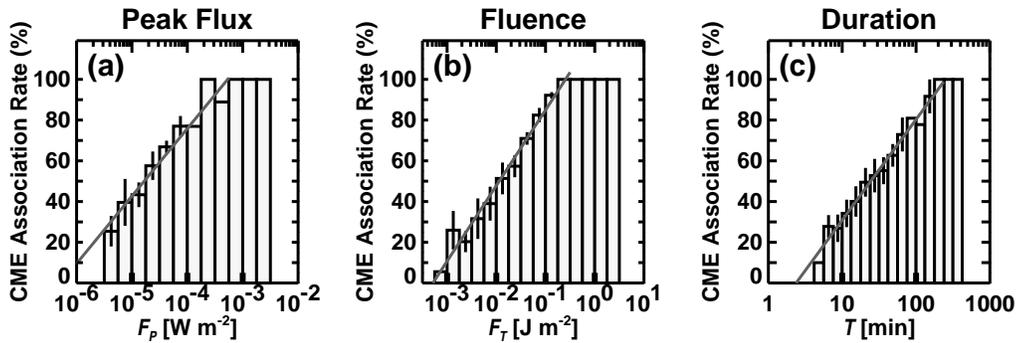}
\caption{CME association rate as a function of (a) X-ray peak flux,
(b) fluence, and (c) duration. The gray straight line is the
least-squares fit to the data points.}
\end{figure*}

\section{Data Set}

The basic flare parameters such as peak flux, fluence, and duration
are available in the Solar Geophysical Data (SGD) and online Solar
Event
Reports\footnote{\url{http://www.sec.noaa.gov/ftpmenu/indices.html}}
provided by NOAA.  The peak flux, measured in the 0.1 - 0.8 nm
wavelength band, determines the rank of X-ray flares. The letters (A,
B, C, M, X) designate the order of magnitude of the peak flux
($10^{-8}$, $10^{-7}$, $10^{-6}$, $10^{-5}$, $10^{-4}$ W~m$^{-2}$,
respectively). The number following the letter is the multiplicative
factor. For example, an M3.2 flare indicates an X-ray peak flux of
$3.2\times10^{-5}$ W~m$^{-2}$. The fluence (total flux) of a flare is
obtained by integrating the X-ray flux in the 0.1 - 0.8 nm band from
its start to end. No background subtraction is applied for the peak
flux and fluence. The flare start time is identified as the first
minute in a sequence of 4 minutes of steep monotonic increase in 0.1 -
0.8 nm flux. The end time corresponds to the time when the flux decays
to a point halfway between the maximum flux and the pre-flare
background level. More than 20,000 flares have been recorded from 1996
to 2005, but not all events were used in this study. We excluded
flares below C3.2 level, since it is very difficult to examine their
CME association. In this paper, a C-class flare means the peak flux is
between C3.2 and C9.9 level.

We used CME data routinely obtained by the C2 and C3 telescopes of the
Large Angle and Spectrometric Coronagraph \citep[LASCO;][]{bruec95} on
board SOHO. We excluded flares corresponding to SOHO/LASCO
downtimes. For the CME occurrence rate studies, usually a 3-hour
criterion is used to define LASCO downtimes \citep{stcyr00,gopal04b},
but we applied a harder criterion in this study. We required at least
two LASCO C2 images were obtained between 0 - 2 hours after the flare
onset.  Examining the CME visibility (detection efficiency) of the
LASCO coronagraphs, \citet{yashi05} found that about half of disk CMEs
associated with C-class flares and $\sim$16\% of disk CMEs associated
with M-class flares were invisible to LASCO, while all CMEs associated
with X-class flares were visible to LASCO. In order to separate the
flares with CMEs from those without CMEs as accurately as possible, we
eliminated C-class flares with longitudes $<60^\circ$ and M-class
flares with longitudes $<30^\circ$. We also eliminated flares at
longitudes $>85^\circ$ because of the possible partial occultation of
the X-ray source, resulting in an underestimate of the X-ray flux.
Thus we used the longitude range $0^\circ-85^\circ$ for X-class
flares, $30^\circ-85^\circ$ for M-class flares, and
$60^\circ-85^\circ$ for C-class flares. There are 5890 flares (above
C3.2 level) listed in SGD, but the locations are not listed for
$\sim$1800 of them.  For the X- and M-class flares, we identified
their locations using solar disk images obtained in X-ray, EUV,
H$\alpha$, and microwave. For C-class flares, we used only those
flares with their locations listed in SGD. Applying all the above
criteria resulted in 98 X-class, and 692 M-class and 575 C-class
flares during the study period.

\begin{figure*}
\epsscale{0.9}
\plotone{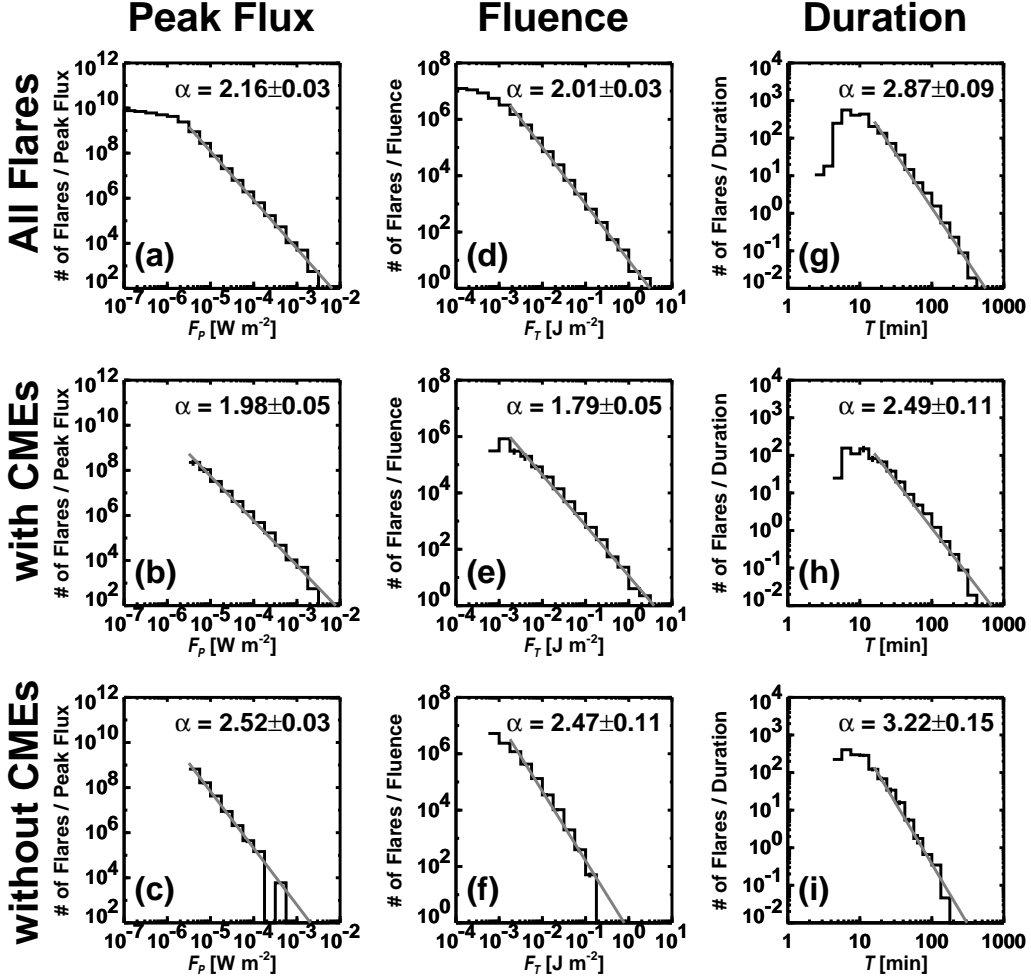}
\caption{Flare frequency distributions as a function of peak flux
(left), fluence (center), and duration (right) for all flares (top),
flares with CMEs (middle), and flares without CMEs (bottom),
respectively. The power-law index~$\alpha$ of each distribution is
shown in the panel. Flares without CMEs have steeper power-law indices
compared to those with CMEs.}
\end{figure*}

\section{CME Associations}

In order to determine the CME association of flares, we used the
SOHO/LASCO CME
Catalog\footnote{\url{http://cdaw.gsfc.nasa.gov/CME\_list/index.html}}
\citep{yashi04} to find the preliminary CME candidates within a 3-hour
time window (90 min before and 90 min after the onset of X-ray
flares). When no candidates were available in the time range, we
checked the original LASCO movies to find any unlisted CMEs in the CME
catalog. If no CME could be observed due to low quality LASCO images
contaminated by solar energetic particles, we excluded the events from
the analysis. The consistency of the association between the flare and
CME candidates was examined by viewing both flare and CME
movies. Eruptive surface signatures, such as filament eruptions and
coronal dimmings, helped ascertain the associations. However, in some
cases, we could not determine with confidence whether their
association was true or false because some flares had obscure eruptive
signatures.  In this case, we abandoned the events to give a clear
true or false answer of the flare's CME associations, and left them as
ambiguous associations. This way, we classified all the flares into
three categories: flares with definite CME association, flares with
uncertain CME association, and flares that definitely lacked CMEs.

Figure~1 shows the CME association rate as a function of X-ray peak
flux (a), fluence (b), and duration (c). The CME association rate has
an error range obtained from the uncertain flare-CME pairs. Assuming
that all of the uncertain events were false, the lower limit of the
CME association was determined by dividing the number of definitive
events by the total number of flares. Similarly we obtained an upper
limit by assuming that all uncertain events were true. We used the
middle of the lower and upper limit as the representative association
rate. This is equivalent to assuming that half of the uncertain events
had true association.

The CME association rate of X-ray flares clearly increased with their
peak flux (Fig~1a). The irregular plot around the X3.0 value
($10^{-3.5}$ W~m$^{-2}$) was due to a small sample size. Only a single
flare without CME association reduced the CME association rate from
100\% to 89\%. The gray line shows the first-order polynomial fit
[$R=33.2\times(\log F_P+6.3)$, where $R$ is the CME association rate
in percentage and $F_P$ is the peak X-ray flux in W~m$^{-2}$].  Note
that this equation is invalid for $R\sim0$. The fit indicates that the
CME association rate will be zero below B5 flares, but there are
observations that B5 or weaker flares have associated CMEs (Gopalswamy
and Hammer, in preparation).

Figure~1b shows a clear increase of CME association rate with the
fluence. In our data set, all X-ray flares with fluence $\ge 0.18$
J~m$^{-2}$ had associated CMEs. Using the least-squares fitting, we
obtained $R=37.1\times(\log F_T +3.3)$, where $F_T$ is the fluence in
J~m$^{-2}$. Again, this equation is invalid for $R \sim 0$.

Figure~1c shows that the CME association rate clearly increased with
flare duration. This confirms the well-known fact that long duration
(or decay) events (LDEs) are likely to be associated with CMEs
\citep{sheel83,kay03}.  In our data set, all X-ray flares with duration
$> 180$ min had associated CMEs. Note that this critical duration (180
min) will change if we use different definitions for flare start and
end times.  We obtain $R=49.4\times(\log T -0.4)$, where $T$ is the
duration in min. The fit indicates that the CME association rate will
be zero at a flare duration of 2.5 min. This equation may be
unreliable for $R \sim 0$, since the definition of flare start and end
may not be good for short-duration flares. It must be noted that the
twenty thousand X-ray flares recorded in SGD from 1996 to 2005, only
31 (0.15\%) flares had their duration $<$ 3 min.

\section{Flare Frequency Distributions}

There were 5890 X-ray flares ($>$ C3.2 level) from 1996 to 2005, but
we could determine the CME associations for selected flares only.
Since different selection criteria were applied for C-, M-, and
X-class flares (see Section~2), we were not able to examine the flare
frequency distributions properly from the selected flares. Therefore
we included the deselected flares, assuming that the CME association
rates of the deselected flares are the same as those of selected
flares. The number of flares with CMEs ($N_{WC}$) in a bin of Figure~1
was calculated from total flare number ($N_{TOT}$) in the same bin
multiplied by CME association rate ($R$): $N_{WC}=N_{TOT} \times
R$. For example, there were 743 flares between the values M1.0 and
M1.8 ($10^{-5.00} \le F_P < 10^{-4.75}$ W~m$^{-2}$) and the CME
association rate of this range was $44.1\pm6.4$\%. Then we estimated
that the numbers of flares with and without CMEs in this range were
$327.7 \pm 47.6$ and $415.3 \pm 47.6$, respectively. We carried out
the same calculation for all the bins in Figure~1, and then obtained
the number of flares with and without CMEs.

\citet{veron02} examined almost 50,000 X-ray flares recorded during
1976 to 2000 and obtained $\alpha = 2.11 \pm 0.13$ for the peak flux,
$\alpha = 2.03 \pm 0.09$ for the fluence, and $\alpha = 2.93 \pm 0.12$
for the duration. First we examined frequency distributions for all
flares to compare them with Veronig et~al.'s results. The top panels
of Figure~2 are frequency distributions as a function of the peak flux
(2a), the fluence (2d), and the duration (2g), showing that all the
three distributions are represented by power-laws. Using the
least-squares method, we obtained a power-law index $\alpha = 2.16 \pm
0.03$ for the peak flux, $\alpha = 2.01 \pm 0.03$ for the fluence, and
$\alpha = 2.87 \pm 0.09$ for the duration. The three power-law indices
are consistent with the results of Veronig et~al. within the error
ranges.

The different frequency distributions for flares with and without CMEs
are shown in the middle and bottom panels of Figure~2. Note that the
error bars are comparable to (or smaller than) the thickness of
plotted lines. The left, center, and right panels show the peak flux
(2b and 2c), the fluence (2e and 2f), and the duration (2h and 2i),
respectively. The distributions are represented by a single power-law
with the different power-law indices (shown in each panel). For flares
with (without) CMEs, we obtained the power-law index $\alpha = 1.98
\pm 0.05$ ($\alpha = 2.52 \pm 0.03$) for the peak flux, $\alpha = 1.79
\pm 0.05$ ($\alpha = 2.47 \pm 0.11$) for the fluence, and $\alpha =
2.49 \pm 0.11$ ($\alpha = 3.22 \pm 0.15$) for the duration. The
power-law distributions of all three parameters are steeper for flares
without CMEs than those for flares with CMEs.

If flares with and without CMEs have different power-law indices, then
combined set of flares should show a double power-law.  However, we
cannot see any indications of a double power-law in Figures~2a, 2d,
and 2g, because flares with CMEs are dominant in the major
ranges. Figure~1a shows that the numbers of flares with and without
CMEs are comparable between C5.7 ($10^{-5.25}$ W~m$^{-2}$) and M3.2
($10^{-4.50}$ W~m$^{-2}$) levels (the CME association rate is from
40\% - 60\% in this range). Flares without CMEs are dominant below
C5.7. However, a significant number of small flares were not detected
due to the high X-ray background (during solar maximum, the X-ray
background reached M level).  Thus, we do not have enough bins to
recognize the double power-law distribution.

\section{Discussion and Conclusions}

Since small flares are unlikely to be associated with CMEs, a
power-law index obtained from small flares should be similar to that
from flares without CMEs. \citet{kruck98} examined the distribution of
small flares in the quiet regions observed by SOHO/EIT, and found the
power-law index to be $2.3 - 2.6$, which is consistent with our
result.  However, from X-ray data, \citet{shimi95} found the index to
be in the range $1.5 - 1.6$. He examined the distribution of the
transient brightenings in active regions observed by Yohkoh/SXT. A
similar power-law index ($1.7 - 1.8$) was found for transient
brightenings in the \ion{Fe}{19} line observed by SOHO/SUMER
\citep{wang06}. The transient brightenings correspond the GOES B-class
flares and below, but the obtained power-law indices were very
different from ours. The different temperature response of the three
instruments above might have resulted in the different power-law
indices. Since the power-law index for the smaller flares is not
observationally determined yet, more studies are needed before
reaching firm conclusions.

\citet{hudso91} showed that smaller flares are able to contribute
dominantly to coronal heating when the power-law index~$\alpha$ is
larger than 2. By separating flares with and without CMEs, we showed
that the flare frequency distribution may obey a double power-law
distribution.  Flares without CMEs dominate at small flare sizes and
with $\alpha = 2.47 \pm 0.11$ for fluence, indicating that nanoflares
contribute to coronal heating if the frequency distribution keeps the
same power-law below the observational limit.

Flares without CMEs are thus a potential source for heating the corona
since they do not have energy loss due to CMEs. The CME kinetic energy
ranges from $10^{28}$ to $> 10^{32}$ erg \citep{gopal04a}, which is
generally higher than the flare energy. In flares with CMEs, more than
half of the released energy is used by CMEs to escape from the Sun. On
the other hand, lack of CMEs allows the entire released energy to go
into flare thermal energy. This is consistent with the observational
result that, for a given flare class, flares without CMEs tend to have
a higher temperature than those with CMEs \citep{kay03}.

CME observations by SOHO/LASCO over the past 10 years enabled us to
perfom an extensive statistical analysis of flares with and without
CMEs. We examined the CME associations of flares from 1996 to 2005
and found that the CME association rate clearly increases with flare's
peak flux, fluence, and duration. These results have been known from
the SMM and Solwind era, but the large sample in our study has shown
these relations clearer. The primary result of this paper is that the
power-law index for the distributions of flares without CMEs is much
steeper than that for distributions of flares with CMEs. This result
supports the possibility that flares without CMEs is a likely source
of coronal heating and is consistent with the observation that flares
without CMEs have a higher temperature.

\acknowledgments

S. Yashiro thanks S. Petty for proofreading. SOHO is a project of
international cooperation between ESA and NASA. The LASCO data used
here are produced by a consortium of the Naval Research Laboratory
(USA), Max-Planck-Institut fuer Aeronomie (Germany), Laboratoire
d'Astronomie (France), and the University of Birmingham (UK). Part of
this effort was supported by NASA (NNG05GR03G).


\end{document}